\documentclass[runningheads,a4paper]{llncs}
\usepackage{graphicx}
\usepackage{booktabs} 
\usepackage{hyperref}
\usepackage{url}
\usepackage{multirow}
\usepackage{amsmath}
\usepackage{cleveref}
\usepackage{float}
\usepackage{balance}
\usepackage{microtype}
\usepackage{comment}
\usepackage{color}
\usepackage{graphics,graphicx}
\usepackage{epsfig}
\usepackage{epstopdf}
\usepackage{colortbl}
\usepackage{caption}
\usepackage{soul}
\usepackage{subcaption}
\usepackage{diagbox}
\usepackage{comment}
\usepackage{cite}
\usepackage{siunitx}
\usepackage[justification=centering]{caption}
\usepackage{enumitem}

\newcommand{\anon}[1]{\textcolor{black}{~(anon.)}}
\newcommand{\anoncite}[1]{\textcolor{black}{[anon.]}}

\begin{document}

\title{Ranking Heterogeneous Search Result Pages using the Interactive Probability Ranking Principle}
\titlerunning{Ranking Heterogeneous SERPs using the iPRP}

%
\author{Kanaad Pathak\inst{}\orcidID{0000-0002-1246-8685} \and
Leif Azzopardi\inst{}\orcidID{0000-0002-6900-0557} \and
Martin Halvey\inst{}\orcidID{0000-0001-6387-8679}}
\authorrunning{K. Pathak et al.}
%
\institute{University of Strathclyde, Glasgow, UK
\email{\{kanaad.pathak,leif.azzopardi,martin.halvey\}@strath.ac.uk}}
\maketitle              
\begin{abstract}
The Probability Ranking Principle (PRP) ranks search results based on their expected utility derived solely from document contents, often overlooking the nuances of presentation and user interaction. 
However, with the evolution of Search Engine Result Pages (SERPs), now comprising a variety of result cards, the manner in which these results are presented is pivotal in influencing user engagement and satisfaction.
This shift prompts the question: How does the PRP and its user-centric counterpart, the Interactive Probability Ranking Principle (iPRP), compare in the context of these heterogeneous SERPs? 
Our study draws a comparison between the PRP and the iPRP, revealing significant differences in their output. The iPRP, accounting for item-specific costs and interaction probabilities to determine the ``Expected Perceived Utility" (EPU), yields different result orderings compared to the PRP.
We evaluate the effect of the EPU on the ordering of results by observing changes in the ranking within a heterogeneous SERP compared to the traditional ``ten blue links''. 
We find that changing the presentation affects the ranking of items according to the (iPRP) by up to 48\% (with respect to DCG, TBG and RBO) in ad-hoc search tasks on the TREC WaPo Collection.
This work suggests that the iPRP should be employed when ranking heterogeneous SERPs to provide a user-centric ranking that adapts the ordering based on the presentation and user engagement.

\end{abstract}

\section{Introduction}

The primary aim of search engines is to help users find information relevant to their specific needs. This process often involves issuing multiple queries, scanning numerous documents, and evaluating the relevance of retrieved documents~\cite{Maxwell2017AExperience}.
To enhance this search experience, results should be presented in a manner that allows users to efficiently identify relevant information~\cite{Joho2006AWeb}. Traditional result cards, consisting of a title, link, and summary, have been designed with this goal in mind. However, today's SERPs feature a variety of card types, including images, data, and suggestions, among others~\cite{Bota2016PlayingWorkload,Azzopardi2018MeasuringMeasure}.
Existing user-centric research suggests that both the design of result cards and the layout of SERPs significantly influence user interactions and their satisfaction and success in search tasks \cite{Chierichetti2011OptimizingPresentation,Wang2016BeyondPresentation,Cutrell2007WhatSearch}.

\begin{figure}[h]
\centering
\includegraphics[width=\textwidth]{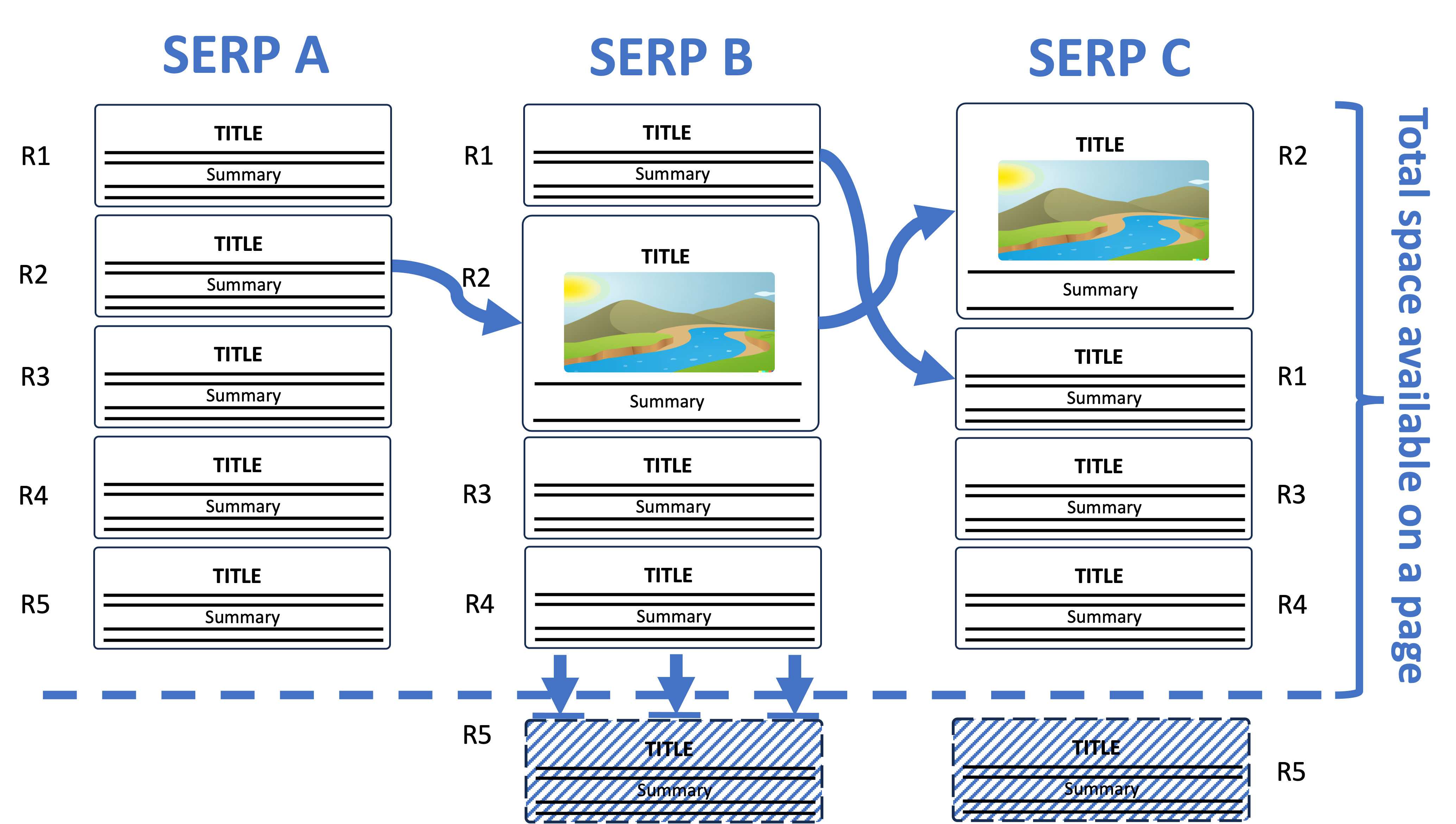}
\caption{
Compared to SERP A, only four cards can be shown above the fold (dotted horizontal blue line) on SERP B and C. However, changing the card type (e.g., TS to TIS) may also lead to changes in the ranking under the iPRP.
}
\label{fig:res_card_ordering}
\end{figure}

For example, in Figure \ref{fig:res_card_ordering}, we can see three different SERP layouts: A, B, and C. In SERP A, all results  \{R1, R2, R3, R4 and R5\} are presented using a title and summary (TS). 
So changing R2 to be presented with a title, image, and summary (TIS) means that it may:
\begin{itemize}
\item attract more (or less) clicks, thus changing its interaction probabilities,
\item take more (or less) time for a user to decide if they want to click the result, or skip over it, thus changing its cost of interaction, and,
\item occupy more (or less) screen space, resulting in a different number of results being displayed above the fold.
\end{itemize}

The Probability Ranking Principle (PRP) ranks results based purely on the decreasing order of their relevance~\cite{Robertson1977TheIr}. Whereas, the Interactive Probability Ranking Principle (iPRP) incorporates interaction probabilities and the cost of processing each result card~\cite{Fuhr2008ARetrieval}. Such considerations might cause R2 to rank higher than R1 in terms of ``Expected Perceived Utility" (EPU) under the iPRP, as demonstrated in SERP C.
Moreover, the type of result cards can significantly influence the overall utility presented to users. Due to space constraints, different card types can alter the number of results displayed on the results page or above the fold, as exemplified by the 5 results in SERP A versus the 4 in SERP B and C. Consequently, adjusting the combination and type of result cards within a SERP introduces trade-offs between EPU, overall utility, and the number of results shown.
In this study, we delve into the potential variations in rankings and performance outcomes when using the iPRP, especially within heterogeneous SERPs that feature diverse result card types. Our investigation is guided by the following Research Questions (RQs):

\begin{itemize}
\item RQ1: What is the impact of different result cards on user behaviour?
\item RQ2: How do the rankings from iPRP in heterogeneous SERPs contrast with those generated by the PRP?
\end{itemize}

\section{Background}
\label{background}
The Probability Ranking Principle (PRP) posits that results should be ranked in descending order of their relevance probability \cite{Robertson1977TheIr}. 
Yet, it makes broad assumptions: that users will systematically inspect each item and that every item demands an identical interaction cost. 
To extend the PRP for interactive scenarios in information retrieval (IR), \cite{Fuhr2008ARetrieval} introduced the interactive PRP (iPRP), considering both costs and benefits of interacting with various result presentations.

Result cards, characterized by varying combinations such as,  titles, images, URLs, and summaries, dictate distinct interaction costs and processing times for users. A substantial body of empirical research has delved into these aspects. 
For instance, while \cite{Rele2005UsingInterfaces} found negligible differences in processing times between list and tabular displays featuring titles and summaries, \cite{Teevan2009VisualRevisitation} observed consistent interaction metrics, such as page clicks, across different card types. 
Yet, variations in user satisfaction were significant. The accuracy of decisions was enhanced when users were presented results with titles complemented by images \cite{Dziadosz2002DoResults}. 
Meanwhile, \cite{Joho2006AWeb} determined that user satisfaction remained stable across the different search result cards, although the efficacy of summaries was notably task-specific. 
Direct comparisons between title-only and title + summary displays revealed that the latter had a pronounced positive effect on relevance assessments \cite{Tombros1998AdvantagesRetrieval}. 
Additionally, the length of the summary emerged as a key factor in task performance. 
While short summaries aligned better with navigational tasks, longer summaries were found to be more beneficial for informational tasks \cite{Cutrell2007WhatSearch,Maxwell2017AExperience}.

These observations highlight a crucial point: the way results are displayed significantly affects how users interact with them and their satisfaction levels. With the ongoing shift from the traditional ``\textit{ten blue links}" approach to more heterogeneous SERPs, an interesting question emerges: How do different card formats affect the way results are ranked? 

The iPRP theorises that such differences will lead to different results rankings. One such instantiation of the iPRP is the Card Model~\cite{Zhang2015InformationInterface} which conceptualizes the interaction process as a cooperative game between two participants: the system and the user. The goal of the game is to maximize the information gain, while trying to minimize the user effort. The model estimates the utility of a displayed card by considering both its presentation cost and the resulting user benefit (which we will refer to as the expected perceived utility).

Consider a given result item \( \mathbf{i} \) and a specified result card “\( \mathbf{card} \)”. In this context, the user can perform actions, denoted as \( \mathbf{A_{i,j}} \), within an action space \( \mathbf{\mathcal{A}} \). 

Here, \( j \) signifies the type of action: for example clicking, skipping scrolling etc.

Additionally \( \mathbf{R_i} \), signifies the relevance of the result item, for example in a relevance space \( \mathcal{R} \) of graded relevance, this can be relevant, partially relevant and non-relevant.
These actions allow the user to transition to subsequent sets of choices within the retrieved results.
Each action \( \mathbf{A_{i,j}} \) undertaken by the user within this space comes with an associated expected benefit \( \mathbf{B(A_{i,j})} \) and a corresponding expected cost \( \mathbf{C(A_{i,j})} \), incurred from performing that specific action, considering the relevance \( \mathbf{R_i} \) of the item. The Expected Perceived Utility (EPU) of a result card, for a result item \( \mathbf{i} \), is thus generally formulated as:

\begin{equation}
\label{card_model_eqn_full}
\begin{aligned}
EPU_{\text{card}}(i) = \sum_{R_i \in \mathcal{R}} \sum_{A_{i,j} \in \mathcal{A}} P(A_{i,j}|R_i)P(R_i)\Big(B(A_{i,j}|R_i)-C(A_{i,j}|R_i)\Big)
\end{aligned}
\end{equation}

This utility can be extended for a result list as $\mathbf{L}$ \cite{Zhang2015InformationInterface}:
\begin{equation}
\label{epu_list}
\begin{array}{ll}
EPU\left( \boldsymbol{L}\right) & = \sum_{i=1}^{n}\left(\prod_{j=1}^{i-1} \big( 1-P(R)_{j}\big)\right) EPU_{\text{card}}(i) \\
\text{subject\ to\ } & 1 \leq W \leq M
\end{array}
\end{equation}

Here, $P(R)_{j}$ represents the relevance probability of the result item, $\mathbf{W}$ is the space occupied by the result card and $\mathbf{M}$ is the total units of screen space available, and $\mathbf{n}$ is the total number of results in the list $L$. 
While the Card Model/iPRP are underpinned by strong theoretical principles, applying them in real-world situations presents challenges related to the estimation of the different costs and benefits. In the following section, we discuss a methodology for estimating the associated costs and benefits, aiming to offer a renewed vision for the Card Model's (specifically the ``Plain Card" model) practical execution and evaluate its influence on result ordering.
\label{sec: method}
\section{Proposed Method}
We begin by considering a common search scenario, where a user is presented with a list items on different result cards. 
The user examines each result card sequentially, choosing either to: (1) click on the result card, or (2) skip the result card and proceed to the next one. 
This is the user model assumed by the interactive Probability Ranking Principle~\cite{Fuhr2008ARetrieval} and implemented in the Card Model~\cite{Zhang2015InformationInterface}. 
In this section, we propose a method to estimate the costs and benefits, to compute the EPU for this setting. We use the following key variables:
\begin{itemize}
    \item $\mathbf{\mathcal{A}}$: Action space, containing actions: clicking, $c$ and skipping, $s$.
    \item $\mathbf{\mathcal{R}}$: Relevance space containing $R$ and $\bar{R}$.
    \item $\mathbf{R_{i}}$: Relevance of an item $i$: relevant, $R$ or non-relevant, $\bar{R}$.
    \item $\mathbf{P(R_{i})}$: Probability of the relevance of item $i$.
    \item $\mathbf{P(A_{i,j}|R_{i})}$: Probability of taking action $A_{i,j}$ given $R_{i}$.
    \item $\mathbf{B(A_{i,j}|R_{i})}$: Benefit of taking action $A_{i,j}$ given  $R_{i}$.
    \item $\mathbf{C(A_{i,j}|R_{i})}$: Cost of taking action $A_{i,j}$ given $R_{i}$.
\end{itemize}

To estimate the EPU, we must first define the costs and benefits associated with each action, given the result card and associated document. In this work, we adopt the suggestion of ~\cite{Fuhr2008ARetrieval}, who proposed using \textit{time} to represent both the benefit (time saved) and the cost (time spent). The rationale is that users invest their time to find relevant result items (a cost), and discovering relevant result item saves them time as they do not need to keep searching for the required information. The time taken for various actions is influenced by the relevance of the result items and the presentation of the result cards.
This is just one method in which benefits can be computed. We can also incorporate other heuristics beyond dwell time such as mouse position and scroll behaviour and incorporate them into our action space for estimating the EPU~\cite{Guo2012BeyondBehavior}, however, we leave that to be incorporated in the future.
Thus, we can calculate the expected cost and benefit based on the summation of the item's relevance, $\mathbf{R}$ (relevant) or $\mathbf{\bar{R}}$ (non-relevant), given an action $\mathbf{A_{i,j}}$. The expected benefit of an action can be written as:
\begin{equation}
\label{benefit}
\mathbf{B(A_{i,j})} 
= P(A_{i,j}|R)B(A_{i,j}|R) + P(A_{i,j}|\bar{R}) B(A_{i,j}|\bar{R})
\end{equation}
\noindent and the expected cost of an action can be written as: 
 
 \begin{equation}
\label{cost}
\mathbf{C(A_{i,j})}
= P(A_{i,j}|R)C(A_{i,j}|R) + P(A_{i,j}|\bar{R}) C(A_{i,j}|\bar{R})
\end{equation}

Given our expressions for the expected cost and benefit, we can re-write the EPU of a card from Equation~\ref{card_model_eqn_full}, where the action space is limited to clicking and skipping and the relevance is binary; for a given item as follows:
\begin{equation}
\label{card_model_eqn}
\begin{aligned}
EPU_{\text{card}}(i) = \sum_{R_i \in \{R, \bar{R}\}} \sum_{A_{i,j} \in \{c, s\}} P(A_{i,j}|R_i)P(R_i)\Big(B(A_{i,j}|R_i)-C(A_{i,j}|R_i)\Big)
\end{aligned}
\end{equation}

An open question is: how to meaningfully estimate these costs and benefits in terms of time? 
The remainder of this paper is divided into several parts to address this. Firstly, we will derive estimations for these variables from the experimental data. Subsequently, we will investigate how different result cards affect the EPU. Lastly, we will evaluate whether incorporating various result cards within the same ranked list has the potential to alter the ranking of results under the card model/iPRP.

\label{sec: methodology}
\section{Experimental Methodology}
\label{sec:design_procedure}
\begin{figure*}
\begin{minipage}{0.45\columnwidth}
  \includegraphics[width=0.9\textwidth, height=4cm]{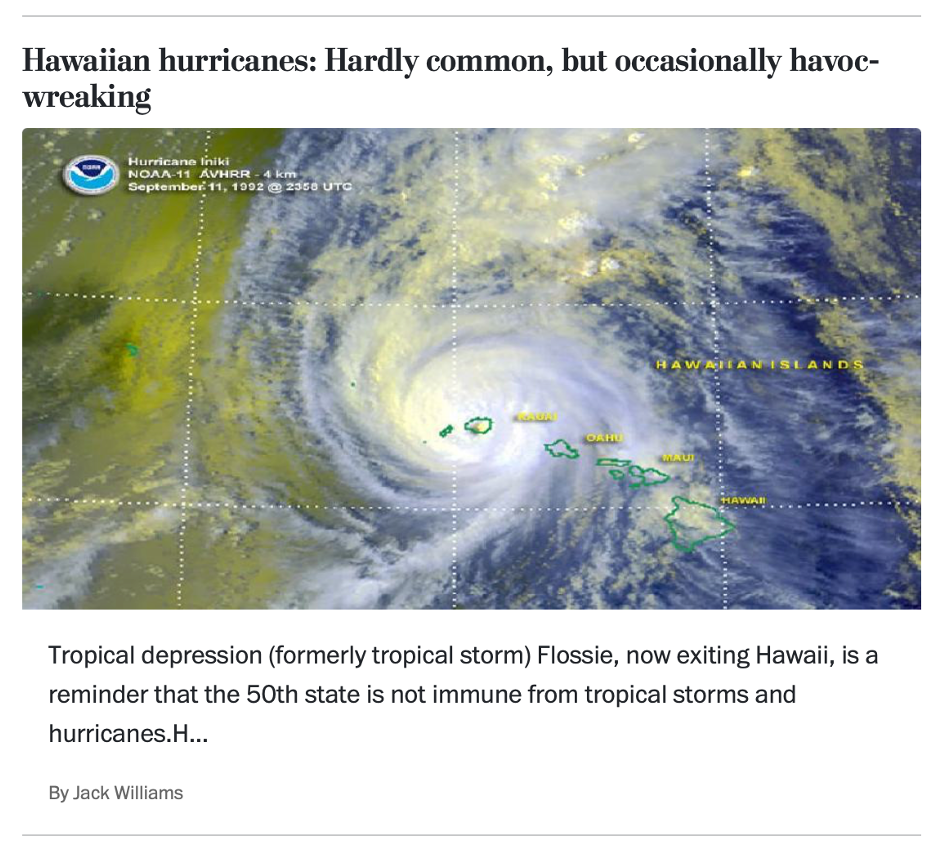}
         \caption{Title+Image+Summary (TIS)}
         \label{fig:ct_2}
\includegraphics[width=0.9\textwidth, height=0.75cm]{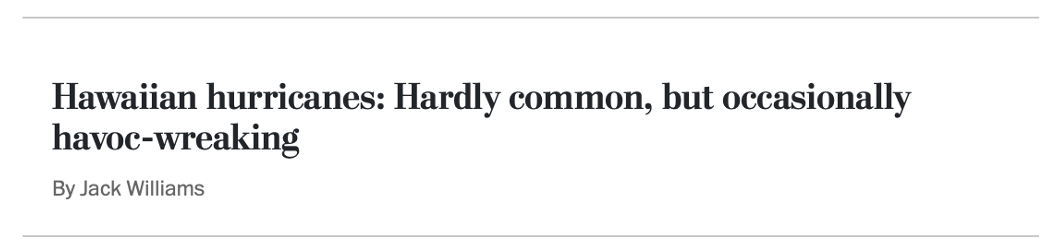}
         \caption{Title Only (T)}
         \label{fig:ct_3}
\end{minipage}
\hfill  
\begin{minipage}{0.45\columnwidth}     
\centering
     \includegraphics[width=0.9\textwidth, height=2.25cm]{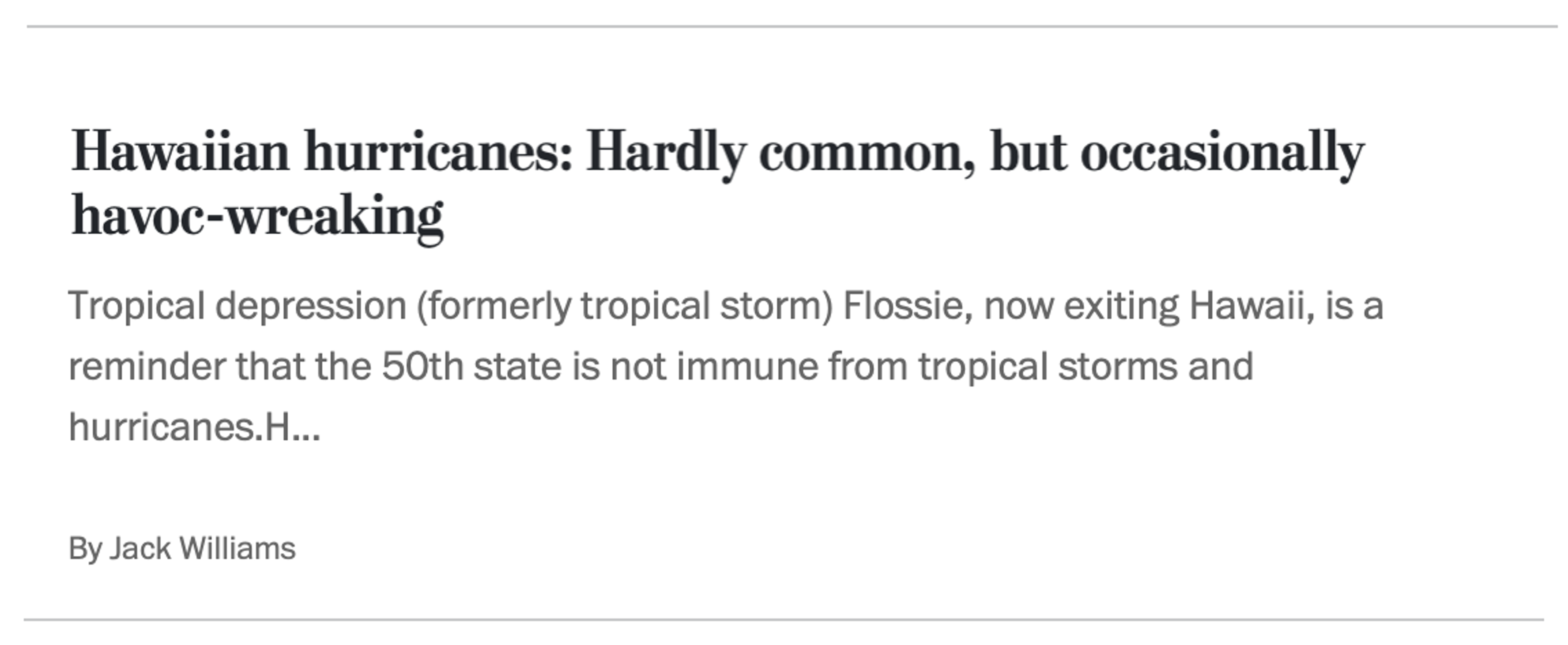}
         \caption{Title+Summary (TS)}
         \label{fig:ct_1}

         \includegraphics[width=0.9\textwidth, height=2.8cm]{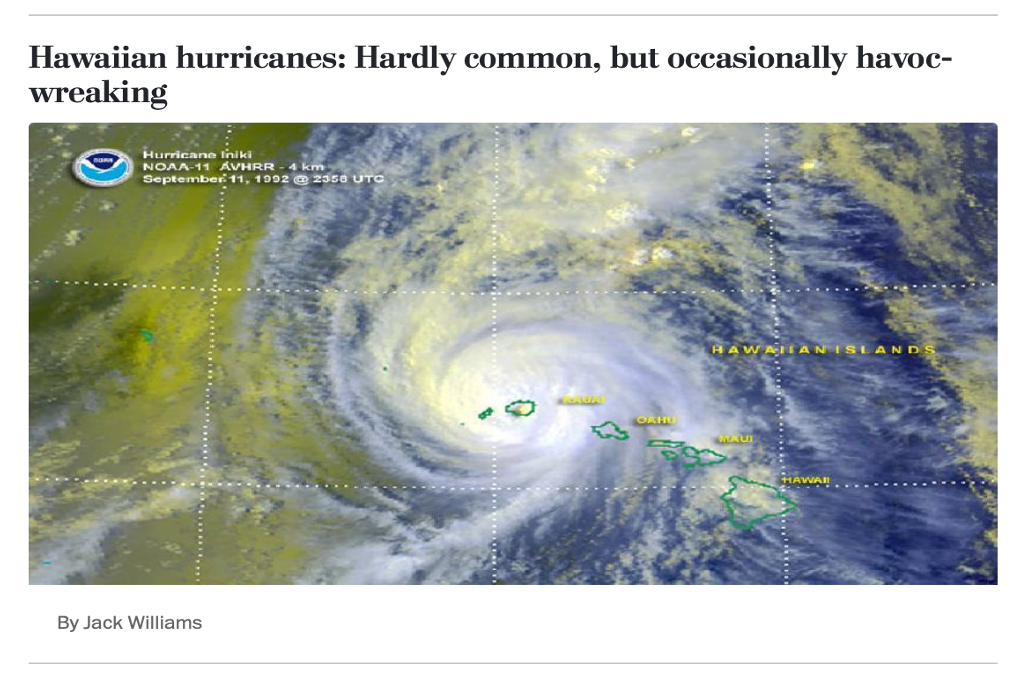}
         \caption{Title+Image (TI)}
         \label{fig:ct_4}
\end{minipage}
\label{fig:card_types}
\end{figure*}

To explore the impact of the iPRP on ranking heterogeneous search engine result pages, we experimented with four different result card types (see Figures [~\ref{fig:ct_2}-\ref{fig:ct_4}]). These cards represent typical variations on SERPs. 

To ground our analysis, we gathered timing data and click data on these result cards across three topics from the TREC WaPo collection, employing 150 annotators. Following this, we utilized the annotation data to estimate interaction probabilities and timing components of EPU, leading to the determination of rankings under the iPRP using the Card Model, as previously detailed.

The four result card types used were:
\begin{enumerate}[label=\textbf({\alph*})] 
\item \textbf{TIS}: These cards displayed the article's Title, Image and Summary (representing the most attractive and informative result card, often used for promoted news articles).
\item \textbf{TI}: These cards displayed the article's Title along with the Image (representing a similar result card to what is used on Google News).
\item \textbf{TS}: These cards displayed the title and summary (representing the default result cards used by the Washington Post).
\item \textbf{T}: These cards only displayed the title of the article (representing the sufficient headlines result card).
\end{enumerate}

To ensure consistency across cards we used the Washington Post's: (1) style sheets to give them all the same look and feel, and (2) summaries which consisted of the lead sentence. The width of the cards was the same (6 columns out of 12 -- using Bootstrap), however, because the different cards housed different elements the heights varied, but in a controlled way, where one TIS card used six rows, one TI card used four rows, one TS card used three rows, and the T card used one row. So, for example, if the SERP had 12 rows of vertical space it could hold either two TIS cards, three TI cards, four TS cards, 12 T cards, or some combination of.

{\bf Collection}\label{sec: collection_used}: This study used the TREC Washington Post Corpus (WaPo) collection from the TREC Common Core 2018 track\footnote{https://trec-core.github.io/2018/}. 
The collection consists of 608,180 news articles and blog posts published between January 2012 and August 2017 categorized into 50 topics for information retrieval tasks. 
We selected three topics for annotation (341: Airport Security, 363: Transportation Tunnel Disaster, and 408: Tropical Storms). These topics were selected because they had at least 100 TREC judgments, of which at least 50 were judged relevant documents, for which there were downloadable images. 
This ensured we had a sufficient mixture of relevant/non-relevant items to annotate and that we could render all card types. The images were downloaded and re-scaled so that images were of the same height and width. 

{\bf Annotations}\label{sec:task_procedure}: An interface was developed to collect annotations. Given a description of the topic, the annotators were shown results styled as one of the different result cards. They were then given the option to click the ``view" button (if they thought it was relevant), or skip the card (if they thought it was unlikely to be relevant). Results were randomly selected without replacement from the pool of TREC documents and result cards were also randomly selected, to minimize any order effects. Participants could annotate up to 50 results (in batches of 10).
We recorded participants' actions (e.g., clicking, skipping) and the time taken to perform these actions.
We recruited 150 participants from the United Kingdom and the United States on Prolific\footnote{https://www.prolific.co}. Participants were required to use a desktop or laptop computer and have English as their first language. 
Pre-screening checks on Prolific ensured they met these criteria. After reading the on-screen information sheet and providing consent, participants completed a set of practice annotations to familiarize themselves with the task and setup. 
Ethics approval (no. 1643) was granted via the departmental ethics committee at the University of Strathclyde for this task, and participants were compensated in line with national working wage guidelines. 
\\
{\bf Participant Demographics}\label{sec:participant_demographics}: 
In total, we collected 6,052 annotations from 150 participants (approx. 40 annotations per participant, approx. 10 annotations per result card per topic per participant.) The study sample comprised a near-equal gender distribution of 77 males, 71 females and two participants preferring not to identify with either gender. Participants ranged in age from 21 to 75 years, capturing a broad spectrum of adult age groups. Within this cohort, a minority of 14 individuals (9.3\%) identified as students, while the majority, 136 participants (90.7\%), were non-students.

A significant proportion of the participants, 83 individuals (55.3\%), were engaged in either full-time or part-time employment. The remaining 67 participants (44.7\%) were not involved in paid employment at the time of the study, which includes groups such as homemakers, retired, or disabled individuals. 

\subsection{Estimating the EPU}
Before we can exactly instantiate the iPRP via the Card Model, we still need to define how we estimate the costs and benefits of clicks and skips, as well as how we estimated the probability of relevance.

For the costs, we used the time spent performing each of the different actions denoted by $T$ i.e.:
\begin{itemize}
\item Cost to click a relevant {\boldmath\(C(A_{i,j}=c|R_{i}=R) = T(c|R)\)} and non-relevant item $\mathbf{C(A_{i,j}=c|R_{i}=\bar{R})= T(c|\bar{R})}$ , and,
\item Cost to skip a relevant item {\boldmath\(C(A_{i,j}=s|R_{i}=R)=T(s|R)\)} and non-relevant item $\mathbf{C(A_{i,j}=s|R_{i}=\bar{R})=T(s|\bar{R})}$.
\end{itemize}

For the benefits, we need to map the gain from a result, to be in the same units as the cost (i.e., in units of time). 
First we assume that users derive no benefit from choosing to ``view" a non-relevant result, or skipping over a result (relevant or not). 

So we can represent the benefit for these as: {\boldmath\(B(A_{i,j}=c|R_{i}=\bar{R})=0\), \(B(A_{i,j}=s|R_{i}=R)=0\)} and {\boldmath\(B(A_{i,j}=s|R_{i}=\bar{R})=0\)} for all $i$ that are not-relevant, i.e. they get no benefit from these actions.
This leaves the final case when a user clicks ``view" on a relevant result. 
We consider that the time spent reading a relevant result $T(read|R)$ facilitates information acquisition, and thus aligns with the concept of time well spent~\cite{Smucker2012Time-basedMeasures}. 

We define our benefit from a relevant click to be the time required to read the result $\mathbf{B(A_{i,j}=c|R) = T(read|R)}$ (we describe below how to estimate this from our measurements)
There are potentially other ways to map the gain of information to time or vice versa (e.g., \cite{Azzopardi2015AnBehavior,Kim2014ModelingSatisfaction}), however, we leave exploring such avenues for future work.

\textbf{Estimation of time: }For each of the card types, given the relevance, we calculated the average time (in seconds) to click or skip the card.

\begin{itemize}
\item $\mathbf{T(c|R)}$ and $\mathbf{T(s|R)}$: For each result presented in the annotation interface, we measured the time from when the result appeared until the user either clicked the ``view" or ``skipped" button next to the result, respectively. We then computed the average time across all results and users for each result card type, taking into account its relevance.
\item $\mathbf{T(read|R)}$: For each relevant result, given a card type; we measured the time spent reading the result. We then recorded the maximum reading time for each card type and user to compute the average reading time across all users for a given card type. This approach ensures that spending less time on a relevant result does not negatively impact the utility value.  For example, if different results are presented in the same card type, depending on the density of the information in the result it may take longer or shorter amounts of time to read it. This could potentially mean that longer reading times would give more benefits. Therefore, if we cap and fix the benefit per card type to the maximum time to read the result, quickly reading a relevant result will not give a small benefit to one card type or vice versa. Thus, in our benefit computation, we account for different reading speeds, lengths and comprehension of information in the result by taking the maximum time to read the result per card type.
\end{itemize}

\textbf{Estimation of probabilities: }To estimate the interaction probabilities, we counted the number of times an item was shown and how often it was clicked or skipped, given its relevance score and card type.  We used maximum likelihood estimation to calculate the probability of each type of interaction occurring for each card type. 

To estimate the probability of relevance in our analysis, we employed the BM25 retrieval function with $\beta=0.75$. Given that BM25 yields an unbounded retrieval score, it was necessary to convert it to a probability. Following the approach from a related study~\cite{Maxwell2017AExperience}, we used a set of previously submitted queries used on this test collection and issued them to our retrieval engine. For every query variation, the top 50 documents were selected. Then, across all the documents retrieved, the BM25 scores were normalized using z-normalization, and subsequently mapped to a range of [0-1] through a logistic curve transformation. A regression model was then constructed, with an $R^2$ value of 0.791, to predict the probability of relevance based on a BM25 score. However, it's worth noting that perfecting this model is not the primary focus of this paper.

Given our estimates of the different components, we calculated the expected perceived utility for a given result list ($\mathbf{EPU(L)}$) (see Equation. \ref{epu_list}) for a list layout using different results card types for each result to determine how the rankings between the iPRP and PRP vary.

\section{Results}
\label{results}
Table \ref{tab:metrics_table} presents the timings, probabilities, and Expected Perceived Utility (EPU) for each card type, with timings measured in seconds. From this data, we crafted a typical user profile, grounding it in the observed timings and interaction probabilities. For every TREC topic, we used the topic title as a query to fetch the top 20 results using BM25 ($\beta=0.75$). We then calculated the EPU values for each card type, which are displayed in the table as reflective of our average user profile. To pinpoint variations among the card types, used one-way ANOVA tests. Post-hoc analysis was done with Tukey's HSD test. In the table, the mean ± standard deviation of the timings and probabilities are shown and significant differences ($p<0.05$) are emphasized using superscripts.

\begin{table}[ht]
\caption{Components of the Utility Function, Probabilities, and Expected Perceived Utility (EPU) for All Card Types. Significant differences in values between the card types are indicated by {a,b,c or d} in superscript. }
\centering
\scriptsize
\begin{tabular}{lllllcccrr}
\toprule
Card Type & \(T(s|\bar{R})\) & \(T(c|\bar{R})\) & \(T(c|R)\) & \(T(s|R)\) & \(P(s|\bar{R})\) & \(P(c|R)\) & \(EPU_{card}\) \\
\midrule
a. TS    & $4.63 \pm 2.42^{c,d}$ & $4.41 \pm 3.31$ & $4.13 \pm 2.32^{d}$ & $5.49 \pm 2.81$ & $0.69 \pm 0.27$ & $0.81 \pm 0.25$ & $11.78 \pm 4.64^{c,d}$ \\
b. TIS   & $4.40 \pm 2.19^{d}$ & $5.15 \pm 3.61^{c,d}$ & $4.38 \pm 2.70^{c,d}$ & $5.86 \pm 3.53^{c,d}$ & $0.73 \pm 0.27$ & $0.82 \pm 0.23$ & $11.77 \pm 4.66^{c,d}$ \\
c. T     & $3.58 \pm 1.62$ & $3.86 \pm 2.42$ & $3.64 \pm 2.05$ & $4.48 \pm 2.23$ & $0.68 \pm 0.30$ & $0.80 \pm 0.25$ & $10.90 \pm 4.14^{d}$ \\
d. TI    & $3.80 \pm 1.67$ & $3.72 \pm 1.95$ & $3.42 \pm 1.64$ & $4.43 \pm 2.21$ & $0.73 \pm 0.25$ & $0.78 \pm 0.26$ & $6.40 \pm 2.89$ \\
\bottomrule
\end{tabular}
\label{tab:metrics_table}
\end{table}

\newpage
\noindent\textbf{RQ1: What is the impact of different result cards on user behaviour?}

Our observations indicate that there are distinct variations in timings and probabilities associated with different card types, and these differences are statistically significant. Referring to Table \ref{tab:metrics_table}, we can see that integrating an image with a T card diminishes its EPU (T: $10.90$, TI: $6.4$). This trend implies that the addition of images can potentially divert or mislead users, thereby compromising their ability to swiftly discern relevant details. The interaction probabilities bolster this argument as they display a decreased probability of interacting with a relevant item when an image is included ($P(c|R)$ for TI: $0.78 \pm 0.26$).

Conversely, augmenting a T card with a summary enhances its EPU (T: $10.9$, TS: $11.78$). Summaries, especially when coupled with titles and images, potentially supply crucial context, enabling users to better assess the accompanying image. They also present an overview of the result's content, which aids users in determining its relevance and deciding about further engagement. Hence, while T cards enriched with summaries do incur a slightly higher processing time (T: $3.71 \pm 1.95$, TS: $4.23 \pm 2.36$), they demonstrate a reduced probability of mistakes, corroborated by the heightened $P(c|R)$ values (T: $0.80 \pm 0.25$, TS: $0.81 \pm 0.25$).

In the context of Expected Perceived Utility (EPU), we discerned variations in the average EPU across card types. Explicitly, TIS and TS cards exhibit a superior EPU compared to T and TI cards. A one-way ANOVA was conducted, unveiling a statistically significant difference in EPU ($F(3, 3996) = 406.33, p < 0.05$). Delving deeper via Tukey's HSD Test, it became evident that TIS, TS, and TI cards possess an EPU surpassing that of T cards. Moreover, TIS and TS cards outperformed TI cards in terms of EPU. However, the distinction between TS and TIS in EPU was not statistically significant (p=0.129).

While our study observed differences in EPU, timings, and interaction probabilities across card types, the probability of a user clicking or skipping a relevant item, intrinsic to their behavior, remained unaffected by the card type. This finding is consistent with \cite{Teevan2009VisualRevisitation}, which reported no variance in click behaviour across diverse interfaces. However, user satisfaction did differ, highlighting individual differences in satisfaction preferences. 
Although the card type does not consistently alter click probabilities across all users, the EPU can encapsulate these individual variations by incorporating additional context such as the time required to process and read items. For instance, even if a particular card type inherently takes longer to process, it could still be more effective for some users due to their personal preferences or cognitive strengths. Such advantages, like lower error rates (higher $P(c|R)$), could counterbalance the longer processing times, resulting in a higher EPU for specific card designs, such as TS cards, for certain users.
Given that these individual card features play a role, an overarching metric, the Expected Perceived Utility (EPU), presents a more holistic view. 
Our findings underscore that the card type significantly affects user interactions with search results. This raises a subsequent question of how mixing card types on a search results page influences the overall rankings, since changing the card type can change its EPU.

\noindent\textbf{RQ2: How do the rankings obtained from heterogeneous SERPs differ compared to the PRP (in terms of performance)?
}
\begin{table}[H]
\centering
\caption{Comparison of RBO, DCG of Page, and TBG for different card type combinations. Results show a statistically significant difference in RBO between different groups of combinations after running a one-way ANOVA of (F(7,31841)=2517.66, $p<0.001$). "$\sim$" shows that there is no statistically significant difference with that row.}
\begin{tabular}{l|lll}
\toprule
\textbf{Combination Type} & \multicolumn{1}{c}{\textbf{RBO}} & \multicolumn{1}{l}{\textbf{DCG of Page}} & \multicolumn{1}{c}{\textbf{TBG of Page}} \\
\midrule
a. Baseline               & $1.000 \pm 0.000$                         & $3.137 \pm 1.625$                             & $3.073 \pm 0.095$ \\
b. T or TI                & $0.952 \pm 0.135^{\sim c}$                & $2.437 \pm 1.405$                             & $1.960 \pm 0.482$ \\
c. TIS or TS              & $0.951 \pm 0.136$                         & $2.437 \pm 1.407^{\sim g,b}$                  & $1.962 \pm 0.478^{\sim b}$ \\
d. TIS or T               & $0.762 \pm 0.251$                         & $2.614 \pm 1.649^{\sim g}$                    & $2.381 \pm 1.130$ \\
e. TS or T                & $0.741 \pm 0.222$                         & $\mathbf{3.588 \pm 2.029}$                             & $\mathbf{4.363 \pm 0.916}$ \\
f. Random                 & $0.637 \pm 0.291$                         & $2.640 \pm 1.646^{\sim d}$                    & $2.413 \pm 0.784^{\sim d}$ \\
g. TS or TI               & $0.505 \pm 0.321^{\sim h}$                & $2.525 \pm 1.636^{\sim b}$                    & $2.318 \pm 0.215$ \\
h. TIS or TI              & $0.501 \pm 0.385$                         & $2.024 \pm 1.384$                             & $1.595 \pm 0.249$ \\
\bottomrule
\end{tabular}
\label{tab: rbo_tab}
\end{table}
To explore the differences between ranking results by EPU and by EU (ordering with the PRP), we ran a simulation using all 50 TREC WaPo topics. In this simulation, we used the EU from retrieved results with BM25 ($\beta = 0.75$) as our baseline. We assumed that the default result card type was 'TS' and that a page could display up to 12 rows. Thereby creating a baseline ranked order similar to ``$n$ blue links". The core of our simulation involved altering this baseline according to the space constraint. Specifically, we selected every result in the list and changed its card type randomly to one of two possibilities, as illustrated in Table \ref{tab: rbo_tab}. For example, the first result might change to TIS, the second to TS, and so on. Since the result page is constrained to 12 rows, a page containing TIS and TS cards can have cards in the following combinations -- TIS, TS, TS or TIS, TIS or TS, TS, TS etc. We repeated this random alteration 100 times for each result list combination type to observe how such changes impacted the ranking order. After applying these changes, we re-ranked the documents in the altered result list in decreasing order of EPU and then compared this new order with our baseline EU ranking. We used the Rank Biased Overlap (RBO) metric~\cite{Webber2010ARankingsb} to measure any changes in ranking order. Additionally, we looked at the DCG and TBG metrics ($h=224$) to see how different SERP layouts affected search result effectiveness.

Our results, presented in Table \ref{tab: rbo_tab}, show that adjusting the presentation of results via different card types to construct heterogeneous SERPs can change document ordering. The RBO metric can quantify this change, however, we acknowledge that RBO is opaque in the sense that it cannot tell us if the change was positive or negative. We leave the exploration of this to future studies that will collect user satisfaction scores to quantify this.

In analyzing DCG scores for our altered result pages, we found that some SERP layouts influenced both RBO and DCG scores similarly. Post-hoc tests using Tukey's HSD Test revealed no significant difference in RBO for certain combinations of card layouts such as T, TI and TIS, TS or TS,TI and TIS or TI. Notably, for DCG, there wasn't a significant difference among several combinations of card layouts, despite the differences in card type mixes.

Time Biased Gain (TBG), accounts for the time spent by the users and their attention on retrieved results~\cite{Smucker2012Time-basedMeasures}. We can observe with the TBG how the costs associated with reading each item in the result list affects the gain of the page. For example the TBG of Page is significantly higher when we combine T cards with TS cards for a result list, where as combining TIS cards with TI cards has a significantly lower TBG compared to the baseline. These results emphasize the role of card types and their arrangement in influencing search result effectiveness and how the time spent assessing the results will affect users' gain. This underscores the need to carefully consider both the presentation and number of search results to optimize user experience (space-utility trafe-off).

Our observations show how the alteration of presentation influences the order of ranked list for the iPRP compared to the PRP. In our implementation of TBG, we have implemented a simplistic user model that assumes linear browsing, like the iPRP. In further work, we aim to explore how changing the presentation affects other complex browsing models.

\section{Discussion \& Future Work}
\label{conclusion}

Our study examined whether the Interactive Probability Ranking Principle (iPRP), instantiated via the Card Model, significantly affects the ranking of search result pages based on the relevance of items and their presentation. We aimed to understand the impact of presentation when ranking heterogeneous result pages with four common types of result cards under the iPRP. We framed the iPRP/Card Model as producing the expected perceived utility of each result presented, factoring in different interaction probabilities and decision-making times for various result card types. This approach contrasts with the original PRP, which only considers item relevance for ranking. Our research focused on two main questions, exploring the Expected Perceived Utility (EPU) of different result card types, and the impact of ranking results by EPU on performance with respect to Rank Biased Overlap (RBO), Discounted Cumulative Gain (DCG) and Time Biased Gain (TBG).

Our findings indicate that in the context of ad-hoc news search for the TREC WaPo dataset, result cards using a title, image, and summary (TIS) or title and summary (TS) yield the highest EPU, which is in line with previous research that finds users tend to be more satisfied with a Title and Summary or a Title and Image~\cite{Rele2005UsingInterfaces,Teevan2009VisualRevisitation,Dziadosz2002DoResults,Joho2006AWeb,Tombros1998AdvantagesRetrieval}. However, these card types also limit the number of cards that can be displayed on the screen, creating a trade-off between space and utility. We found that this trade-off is crucial, as the choice of result card type can significantly affect SERP effectiveness through the re-ordering of documents at higher ranks, as evidenced by RBO, DCG and TBG measurements.

Moreover, we show how altering the result card type on a SERP changes the ranking of items on the SERP (and also the DCG and TBG of the page) compared to a homogeneous result card format. This suggests that when ranking heterogeneous result pages, it may be possible to manipulate the presentation of results to demote or promote items in the ranking, given the differences in how people engage with different card types. This can raise some ethical concerns as manipulating the presentation can be used to bias users toward specific results. Diving into more detail about these ethical considerations is currently out of the scope of this paper. 

In conclusion, our study underscores the importance of considering the presentation of search results when designing ranking algorithms. The perceived relevance of items can change the ranking of documents depending on the presentation of results. We have established that presentation matters when ranking, and that presentation effects can be encoded within a theoretical framework to estimate the expected ``perceived" utility. 

However, due to the highly controlled nature of our study, the applicability of our observations on other domains such as e-commerce or travel search remains to be explored. Other future work will also investigate different ways to estimate the costs and benefits of interaction and different result card types/styles, and how they impact user performance and satisfaction when interacting with heterogeneous search engine result pages. The findings from our study sets up the foundation for instantiating the benefits and costs and estimating the EPU with a simple model for an ad-hoc search task for the TREC WaPo datset. 

Using this new understanding of EPU, we can think about adapting and optimizing heterogenous SERPs according to user preferences. We should thus theoretically be able to create SERP layouts that increase user satisfaction and make the user more efficient at finding relevant information.

\section{Acknowledgements}
We want to thank the reviewers for their insightful suggestions and feedback and all the participants who took part in the study. The work reported here is funded by the DoSSIER project under European Union’s Horizon 2020 research and innovation program, Marie Skłodowska-Curie grant agreement No 860721

\bibliographystyle{splncs04}
\bibliography{9_references}

\begin{thebibliography}{10}
\providecommand{\url}[1]{\texttt{#1}}
\providecommand{\urlprefix}{URL }
\providecommand{\doi}[1]{https://doi.org/#1}

\bibitem{Azzopardi2018MeasuringMeasure}
Azzopardi, L., Thomas, P., Craswell, N.: {Measuring the utility of search engine result pages: An information foraging based measure}. 41st International ACM SIGIR Conference on Research and Development in Information Retrieval, SIGIR 2018 pp. 605--614 (6 2018). \doi{10.1145/3209978.3210027}, \url{https://dl.acm.org/doi/10.1145/3209978.3210027}

\bibitem{Azzopardi2015AnBehavior}
Azzopardi, L., Zuccon, G.: {An analysis of theories of search and search behavior}. In: ICTIR 2015 - Proceedings of the 2015 ACM SIGIR International Conference on the Theory of Information Retrieval (2015). \doi{10.1145/2808194.2809447}

\bibitem{Bota2016PlayingWorkload}
Bota, H., Zhou, K., Jose, J.M.: {Playing your cards right: The effect of entity cards on search behaviour and workload}. CHIIR 2016 - Proceedings of the 2016 ACM Conference on Human Information Interaction and Retrieval pp. 131--140 (3 2016). \doi{10.1145/2854946.2854967}, \url{https://dl.acm.org/doi/10.1145/2854946.2854967}

\bibitem{Chierichetti2011OptimizingPresentation}
Chierichetti, F., Kumar, R., Raghavan, P.: {Optimizing two-dimensional search results presentation}. Proceedings of the 4th ACM International Conference on Web Search and Data Mining, WSDM 2011 pp. 257--266 (2011). \doi{10.1145/1935826.1935873}

\bibitem{Cutrell2007WhatSearch}
Cutrell, E., Guan, Z.: {What are you looking for?: An eye-tracking study of information usage in Web search}. In: Conference on Human Factors in Computing Systems - Proceedings (2007). \doi{10.1145/1240624.1240690}

\bibitem{Dziadosz2002DoResults}
Dziadosz, S., Chandrasekar, R.: {Do thumbnail previews help users make better relevance decisions about web search results?} In: SIGIR Forum (ACM Special Interest Group on Information Retrieval) (2002). \doi{10.1145/564437.564446}

\bibitem{Fuhr2008ARetrieval}
Fuhr, N.: {A probability ranking principle for interactive information retrieval}. Information Retrieval  \textbf{11}(3),  251--265 (6 2008). \doi{10.1007/s10791-008-9045-0}

\bibitem{Guo2012BeyondBehavior}
Guo, Q., Agichtein, E.: {Beyond dwell time: Estimating document relevance from cursor movements and other post-click searcher behavior}. WWW'12 - Proceedings of the 21st Annual Conference on World Wide Web pp. 569--578 (2012). \doi{10.1145/2187836.2187914}, \url{https://dl.acm.org/doi/10.1145/2187836.2187914}

\bibitem{Joho2006AWeb}
Joho, H., Jose, J.M.: {A comparative study of the effectiveness of search result presentation on the Web}. In: Lecture Notes in Computer Science (including subseries Lecture Notes in Artificial Intelligence and Lecture Notes in Bioinformatics). vol. 3936 LNCS (2006). \doi{10.1007/11735106{\_}27}

\bibitem{Kim2014ModelingSatisfaction}
Kim, Y., Hassan, A., White, R.W., Zitouni, I.: {Modeling dwell time to predict click-level satisfaction}. WSDM 2014 - Proceedings of the 7th ACM International Conference on Web Search and Data Mining pp. 193--202 (2014). \doi{10.1145/2556195.2556220}, \url{https://dl.acm.org/doi/10.1145/2556195.2556220}

\bibitem{Maxwell2017AExperience}
Maxwell, D., Azzopardi, L., Moshfeghi, Y.: {A study of snippet length and informativeness behaviour, performance and user experience}. SIGIR 2017 - Proceedings of the 40th International ACM SIGIR Conference on Research and Development in Information Retrieval pp. 135--144 (8 2017). \doi{10.1145/3077136.3080824}

\bibitem{Rele2005UsingInterfaces}
Rele, R.S., Duchowski, A.T.: {Using eye tracking to evaluate alternative search results interfaces}. In: Proceedings of the Human Factors and Ergonomics Society (2005). \doi{10.1177/154193120504901508}

\bibitem{Robertson1977TheIr}
Robertson, S.E.: {The probability ranking principle in ir} (1977). \doi{10.1108/eb026647}

\bibitem{Smucker2012Time-basedMeasures}
Smucker, M.D., Clarke, C.L.A.: {Time-based calibration of effectiveness measures}. SIGIR'12 - Proceedings of the International ACM SIGIR Conference on Research and Development in Information Retrieval pp. 95--104 (2012). \doi{10.1145/2348283.2348300}

\bibitem{Teevan2009VisualRevisitation}
Teevan, J., Cutrell, E., Fisher, D., Drucker, S.M., Ramos, G., Andr{\'{e}}, P., Hu, C.: {Visual snippets: Summarizing web pages for search and revisitation}. In: Conference on Human Factors in Computing Systems - Proceedings (2009). \doi{10.1145/1518701.1519008}

\bibitem{Tombros1998AdvantagesRetrieval}
Tombros, A., Sanderson, M.: {Advantages of query biased summaries in information retrieval}. SIGIR Forum (ACM Special Interest Group on Information Retrieval)  (1998). \doi{10.1145/290941.290947}

\bibitem{Wang2016BeyondPresentation}
Wang, Y., Yin, D., Jie, L., Wang, P., Yamada, M., Chang, Y., Mei, Q.: {Beyond ranking: Optimizing whole-page presentation}. WSDM 2016 - Proceedings of the 9th ACM International Conference on Web Search and Data Mining pp. 103--112 (2 2016). \doi{10.1145/2835776.2835824}, \url{https://app.litmaps.com}

\bibitem{Webber2010ARankingsb}
Webber, W., Moffat, A., Zobel, J.: {A similarity measure for indefinite rankings}. ACM Transactions on Information Systems  \textbf{28}(4) (2010). \doi{10.1145/1852102.1852106}

\bibitem{Zhang2015InformationInterface}
Zhang, Y., Zhai, C.: {Information retrieval as card playing: A formal model for optimizing interactive retrieval interface}. In: SIGIR 2015 - Proceedings of the 38th International ACM SIGIR Conference on Research and Development in Information Retrieval (2015). \doi{10.1145/2766462.2767761}

\end{thebibliography}
\end{document}